\documentclass[fleqn,10pt]{wlscirep}

\usepackage{graphicx,graphics,times,color}
\usepackage{bm}
\usepackage{longtable}
\usepackage{color}
\usepackage{lipsum}
\usepackage{epstopdf}
\usepackage{amsmath}

\mathchardef\minus="002D
\newcommand{\ket}[1]{\left|{#1}\right\rangle}

\begin{document}

\title{Integrated information storage and transfer with a coherent magnetic
device}
\author[1]{Ning Jia}
\author[2]{Leonardo Banchi}
\author[2,*]{Abolfazl Bayat}
\author[1]{Guangjiong Dong}
\author[2]{Sougato Bose}

\affil[1]{State key laboratory of precision spectroscopy, Department of Physics, East
China Normal University, Shanghai 200062, China}
\affil[2]{Department of Physics and Astronomy, University College London, Gower
Street, WC1E 6BT London, United Kingdom}
\affil[*]{abolfazl.bayat@ucl.ac.uk}
\date{\today}

\begin{abstract}
Quantum systems are inherently dissipation-less,
making them excellent candidates even for classical information processing.
We propose to use an array of large-spin quantum magnets for realizing a
device which has two modes of operation: memory and data-bus.
While the weakly interacting low-energy levels are used as memory to store
classical information (bits), the high-energy levels strongly interact with
neighboring magnets and mediate the spatial movement of information through
quantum dynamics.
Despite the fact that memory and data-bus require different features, which
are usually prerogative of different physical systems -- well isolation for
the memory cells, and strong interactions for the transmission -- our proposal
avoids the notorious complexity of hybrid structures.
The proposed mechanism can be realized with different setups. We
specifically show that molecular magnets, as the most promising technology,
can implement hundreds of operations within their coherence time, while
adatoms on surfaces probed by a scanning tunneling microscope is a future
possibility.
\end{abstract}

\maketitle


\section*{Introduction}

{
The ultimate fate of the miniaturization of information processing devices
naturally leads to the quantum regime even for realizing classical
computers.
Although
quantum computation holds the promise to be the next step in the evolution
of information technology, quantum algorithms have been proved to speed up
only very specific computational tasks, notably prime factorization \cite%
{shor1997polynomial} and database search \cite{grover1996fast}.
Moreover, quantum mechanical systems are, in principle, based on
unitary operations which are reversible and thus dissipation-less.
In view of this, using a quantum device might also provide a solution to
certain technological obstacles in classical information technology, {\it
e.g.} heat production.
 }
Such information processing is also less demanding
with respect to quantum coherence and their realization is thus less
challenging.
Additionally the miniaturization of electronics (even in the context of
classical computation) with the demands of more data
and functional density naturally leads to the quantum world.

Any information processing device essentially needs at least
two different units to
operate properly, namely a long-lived memory and a fast data-bus for
communication between different registers or processors.
%
The cells of the memory unit have to be well isolated from the rest of the
system. There are several atomic scale spin
systems which show suitable properties
for operating as memories, such as nitrogen vacancy centers in
diamond \cite{fuchs2011quantum},
nuclear spins in solid state systems \cite{morton2008solid,pla2014coherent},
molecular magnets \cite{bertaina2008quantum}
and adatoms on surfaces
\cite{loth2012bistability,delgado2012storage,khajetoorians2011realizing,miyamachi2013stabilizing}.
In contrast,
for information transfer, a strong interaction between the cells of the
data-bus is required for fast operation within the coherence time. Good
examples of quantum systems with strong interactions includes, ion traps
\cite{kielpinski2002architecture}, superconducting qubits \cite%
{barends2014superconducting}, electronic spins in gated quantum dots \cite%
{petta2005coherent,shulman2012demonstration}, and donors in silicon \cite%
{kane1998silicon,zwanenburg2013silicon}.
The opposite demands for isolated memory and strongly interacting data-bus
units make it notoriously difficult to implement both units in the same
physical device. While hybrid structures
(e.g. atom-photon, superconducting qubits-microwave, nuclear-electronic spin)
have been proposed
\cite{monroe2014large,xiang2013hybrid,carretta2013quantum}
for fulfilling this
task, a very high degree of precision is needed to control two different
physical systems and their interaction, in order to transfer information
from one system to another.

In this work we show that nano-magnets with a large half-integer spin
momentum can simultaneously act both as memory and data-bus for information
transfer within the same setup without the complexity of hybrid structures.
In fact, the spin levels in neighboring sites can interact through different
mechanisms resulting in exchange interactions which may vary several orders
of magnitude between different spin levels. In large-spin systems, we
propose to use the flexibility in selecting two weakly interacting
low-energy spin levels for encoding a classical bit, while strongly
interacting high-energy spin levels, act quantum mechanically for
transferring such information between distant memory cells. 
Information transfer between the two units, namely the high and low energy
subspaces, is achieved via global electromagnetic pulses acting on the whole
system.

We specifically consider an array of high-spin magnets which interact
through the Heisenberg Hamiltonian with a large zero-field energy splitting
\cite{gatteschi2006molecular}. We show that such a large energy-level
separation together with the inherent selection rule determined by the
interaction results in different effective exchange interaction for
the low- and high-energy subspaces. This in turn implies that the low-energy
levels display a weak effective interaction, making them suitable for
storage, while high-energy levels result in a strong effective interaction
which can be exploited for fast information transmission.
Although higher spin systems have been proposed for quantum communication
\cite{romero2007quantum,bayat2007transfer,bayat2014arbitrary,ghosh2014emulating},
neither of them can implement both the memory and data-bus.
Our proposed mechanism can be realized in different physical implementations
of high-spin magnets with large zero-field splitting. This includes magnetic
adatoms on surfaces \cite{rau2014reaching}, donors on silicon \cite%
{kane1998silicon,zwanenburg2013silicon} and molecular magnets \cite%
{troiani2011molecular,timco2009engineering,Nakahara:1320438,ardavan2007will,mitrikas2008electron}%
. We specifically consider the latter as a testbed for implementing our
proposal. Indeed, molecular magnets have recently attracted lot of
attentions thanks to the flexibility in engineering their properties through
chemical synthesis \cite{timco2009engineering,Nakahara:1320438} and their
long-coherence time \cite{ardavan2007will,mitrikas2008electron}. Our
proposal, is fully accessible to current technology and allows for hundreds
of operations using the same parameters achieved in recent experiments
\cite{timco2009engineering,ardavan2007will,mitrikas2008electron}
within the coherence time of the system.

\section*{Introducing the Model}

We consider a one dimensional system composed of $N$ quantum nanomagnets
with a certain spin $S$. The magnetic interaction is described by the
Hamiltonian
\begin{align}
  H_{\mathrm{tot}} &= \sum_{i=1}^N H^{\rm s}_{i} +
  J \sum_{i=1 }^{N-1} \vec S_i\cdot \vec S_{i+1} ~,
\label{Htot}
\end{align}
where $J$ is the strength of the isotropic exchange interaction between
magnets and $H^{\rm s}_{i}$ is the local Hamiltonian acting on the
$i$th magnet. As a paradigmatic model we consider
\begin{align}
  H^{\rm s}_{i} &= \mu_B\vec B\cdot g_i\cdot\vec S_i + D
\left(S_i^z\right)^2 + E \left[(S_i^x)^2- (S_i^y)^2\right] ~,
\label{Hs}
\end{align}
where $\vec B$ is the magnetic field, $g_i$ is the position dependent
Land\'e $g$-factor, $\mu_B$ is the Bohr magneton, $D$ models the zero field
splitting, and $E$ represents the planar anisotropy in the crystal field
interaction. For the moment we consider no applied magnetic field, so $\vec
B=0$.

\begin{figure}[t]
\centering
\includegraphics[width=.45\textwidth]{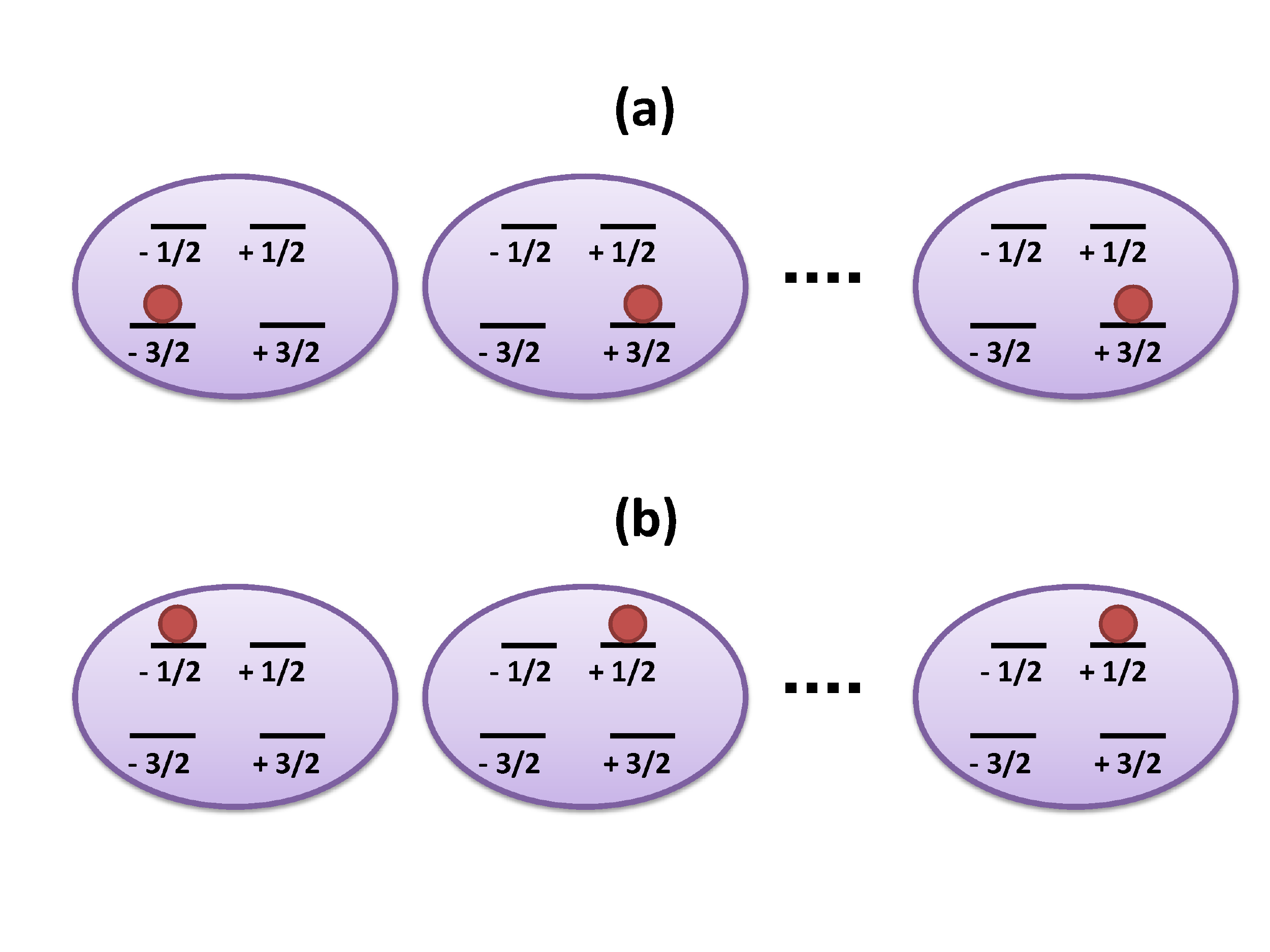}
\caption{Array of $S=3/2$ spins with large negative zero-field splitting $D$
initialized in: (a) the low-energy subspace in the state of $\left|{%
-3/2,+3/2,...,+3/2}\right\rangle$; (b) high-energy subspaces in the state of
$\left|{-1/2,+1/2,...,+1/2}\right\rangle$. The transition between the low
and high energy subspaces can be achieved with a global resonant pulse.}
\label{fig1}
\end{figure}

When $S$ is half integer the eigenstates of \eqref{Hs} comes into pair of
degenerate levels (called Kramers doublet \cite{gatteschi2006molecular})
with opposite magnetization $m$ along the $z$ direction (see Fig.~\ref{fig1}%
).
The states $\left|{m{=}\pm\frac12}\right\rangle$, which are the stable
states when $D>0$, are not suitable to implement a memory because a magnetic
field, whatever small, can induce a transition between them. On the other
hand, the states $\left|{m{=}\pm S}\right\rangle$, which are the stable
states when $D<0$, represent a good candidate to implement a classical bit
in a quantum memory because there is no direct physical coupling between
them. In fact, since a jump between these two states can only occur via
multiple-step processes, bit flip errors are exponentially suppressed.

\section*{Effective dynamics in the low-energy and high-energy subspaces}

\begin{figure}[t]
  \centering
  \includegraphics[width=.5\textwidth]{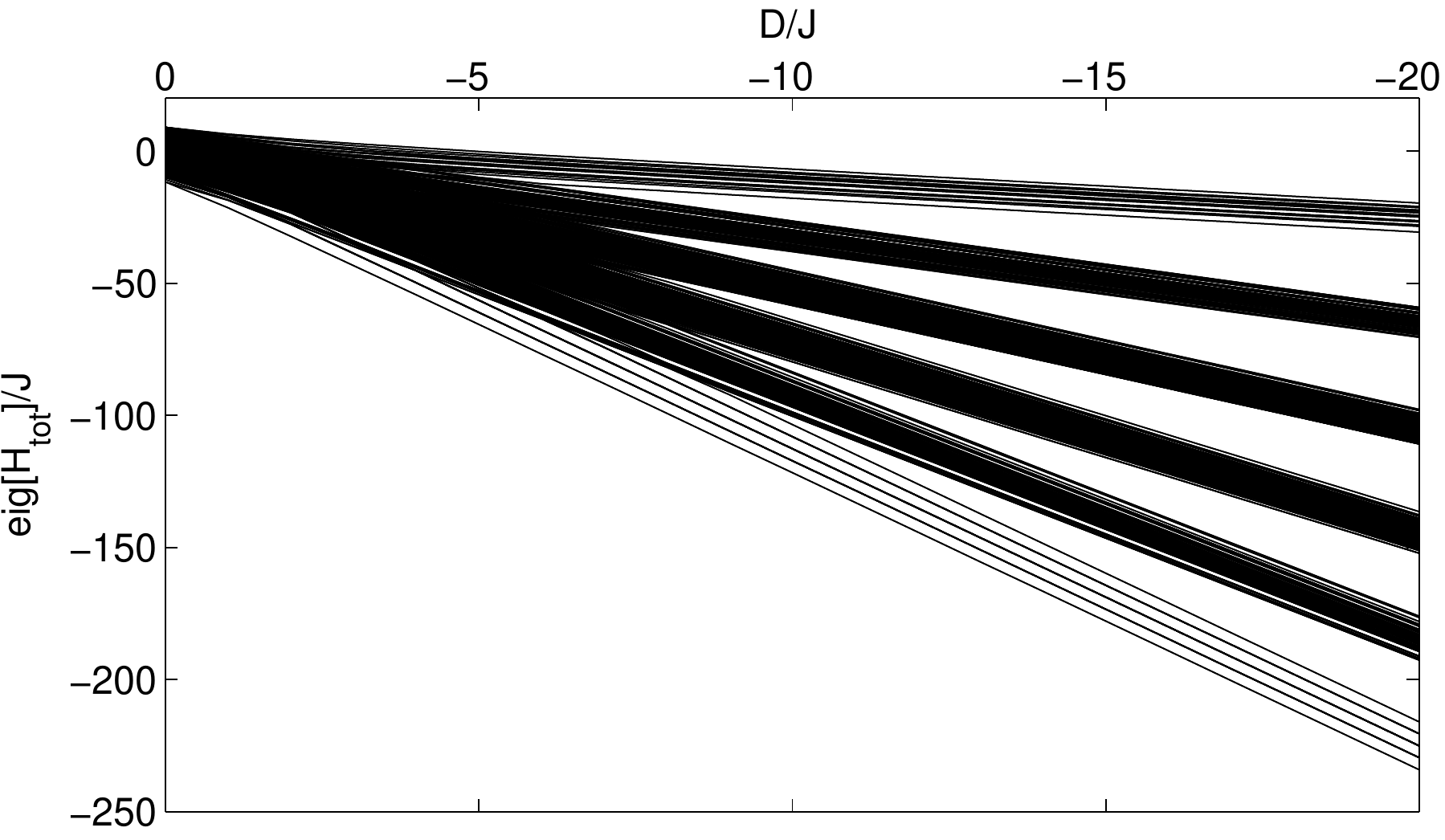}
  \caption{Energy spectrum of $H_{\rm tot}$ for $N{=}5$, $\vec B{=}0$, and
  $E{=}0$, as a function of $D/J$. When $|D|\gg J$ a band structure appears in
  the spectrum. The lowest energy band (composed by the $2^N$ levels in the
  subspace $\{\ket{ {\pm}3/2}\}^{\otimes N}$) forms the memory subspace, while
  the highest energy band (composed by the $2^N$ levels in the
  subspace $\{\ket{ {\pm}1/2}\}^{\otimes N}$) forms the data-bus subspace.
}
  \label{fig:bands}
\end{figure}

In this section we prove that the low-energy Hilbert space (that we call
memory subspace) $\mathcal{H}_{\mathrm{mem}}$ is suitable to store
information, while the high energy Hilbert space $\mathcal{H}_{\mathrm{bus}}$
can be used to implement a data-bus for fast information transfer between
remote memory cells. We start our analysis by deriving two effective
Hamiltonians respectively within the two subspaces. For the moment we
concentrate on $S=3/2$ spin systems
(though later in the paper we will extend our analysis to higher spin
systems) where
\begin{align}
\mathcal{H}_{\mathrm{mem}} &= \{\left|{3/2}\right\rangle,\left|{{-}3/2}%
\right\rangle\}~, & \mathcal{H}_{\mathrm{bus}} &= \{\left|{1/2}%
\right\rangle,\left|{{-}1/2}\right\rangle\}~.  \label{Hilbert32}
\end{align}
{ For negative $D$, in the regime $|D|\gg J$, these two effective subspaces
become energetically well separated. To see this, in Fig.\ref{fig:bands} we
plot the spectrum of $H_{\rm tot}$ as a function of $D$. As it is evident
from the figure, a band structure appears when $|D|\gg J$
in which the lowest band is formed by states in
the memory subspace $\mathcal{H}_{\mathrm{mem}} ^{\otimes N}$,
while the highest band is formed by states in
the data-bus subspace $\mathcal{H}_{\mathrm{bus}} ^{\otimes N}$.
If we initialize our systems in one the bands, throughout the dynamics
other bands are hardly populated. This suggests that there should be an
effective description for the dynamics within the memory and data-bus subspaces.
In the next section we provide effective Hamiltonians for each of these
subspaces.
}


\subsection*{Spin dynamics in the low-energy subspace}

We now consider the regime where $D<0$ and $|D|\gg J$ while $\vec
B=0$. In this regime the states $\left|{{\pm}3/2}\right\rangle$ are
degenerate and well separated from the states $\left|{{\pm}1/2}\right\rangle$%
. This allows us to get an effective interaction between the low energy
states $\left|{{\pm}3/2}\right\rangle$ which is mediated through a
``virtual'' coupling with the high energy states.

We derive the effective Hamiltonian using the theory presented in Supplementary
Material, which is based on two key assumptions: (i) large energy separation
($\approx 2D$)
between the states $\left|{{\pm}3/2}\right\rangle$ and the states $\left|{{%
\pm}1/2}\right\rangle$; (ii) no initial population of the states $\left|{{\pm%
}1/2}\right\rangle$. We found that up to the second order in $J/D$ and $E/D$
one gets
\begin{align}
H_{\mathrm{mem}}^{\mathrm{eff}} = \sum_j &J_{\mathrm{mem}}\left(\tau^x_j
\tau^x_{j+1} +\tau^y_j \tau^y_{j+1}\right) +  \label{HeffSlow} \\
&\Delta_{\mathrm{mem}} \tau_j^z\tau_{j+1}^z +\eta_{\mathrm{mem}}
\tau_j^z\tau_{j+2}^{z} ~,  \notag
\end{align}
where,
\begin{align}
J_{\mathrm{mem}} &= \frac{9J}{64 D^2}(J^2+16 E^2) ~,  \label{JeffSlow} \\
\Delta_{\mathrm{mem}} &= \frac{9 J}{4}\left(1-\frac{J}{8D}\right)
-J\frac{9E^2}{2D^2} - \frac{\xi J^3}{256 D^2}
~, \\
\eta_{\mathrm{mem}} &= \frac{27 J^3}{128 D^2}~,  \label{JDeltaSlow}
\end{align}
and $\xi=90$ in the bulk and $\xi=63$ at the boundaries.
\begin{figure}[t]
\centering
\includegraphics[width=.45\textwidth]{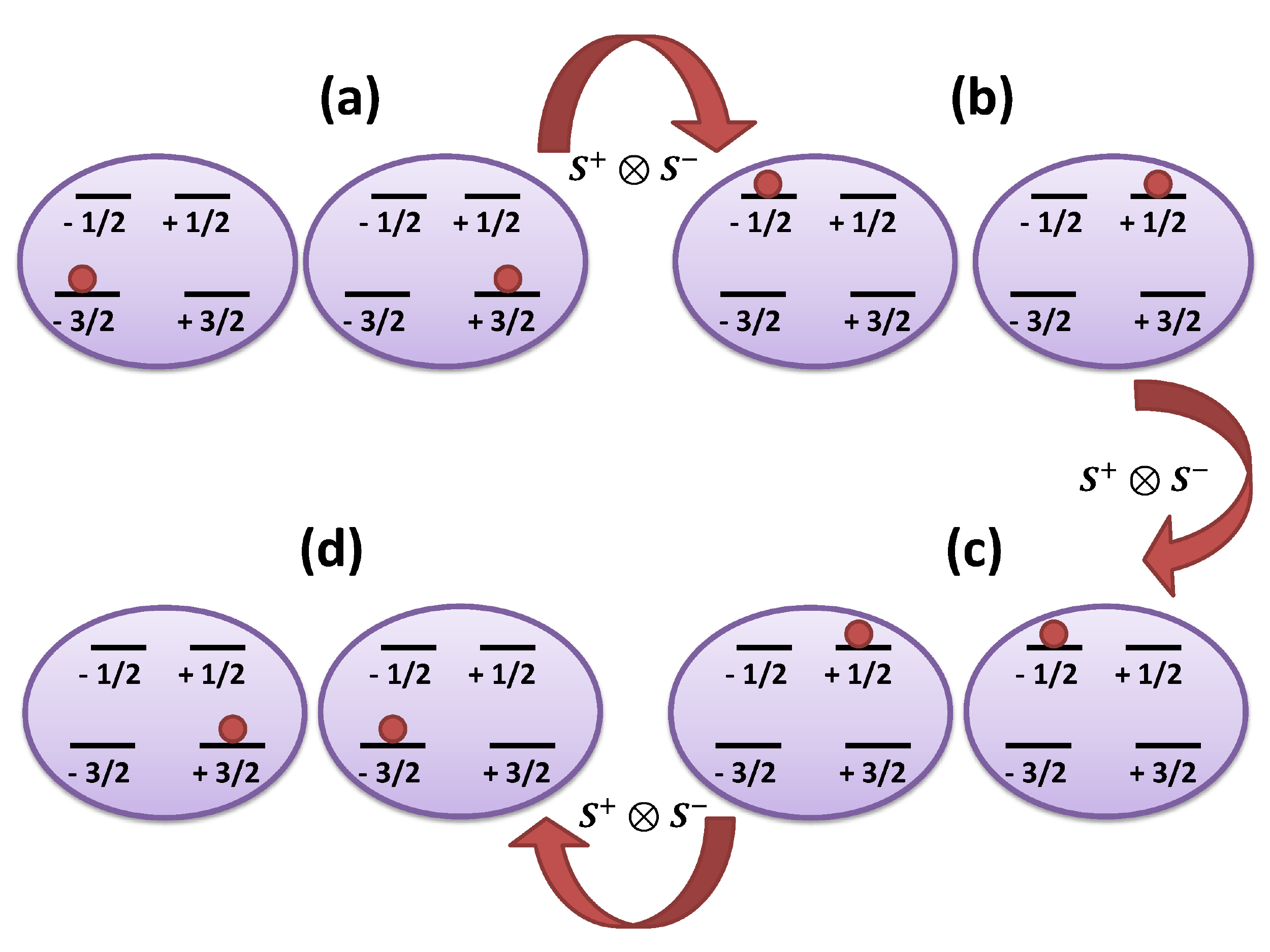}
\caption{The third-order effective hopping Hamiltonian in the low-energy
subspace can be explained by the application of $(S^+\otimes S^-)^3$ (with $%
S^\pm=S^x\pm iS^y$), which arises in the third order perturbation theory
used for getting the effective Hamiltonian (see Supplementary Material for more details).
In fact, the three consecutive operations of the term $S^+\otimes S^-$
result in spin swap in the low-energy subspace through virtually populating
the high-energy states. We show the states
$( S^+\otimes S^-)^n \ket{{-}3/2,3/2}$, for
$n=0$ (a),
$n=1$ (b),
$n=2$ (c),
$n=3$ (d),
which are populated during the process.
}
\label{FigSlow}
\end{figure}
In \eqref{HeffSlow} the matrices $\tau_{x,y,z}$ are Pauli operators defined
in the effective subspace $\{ \left|{{\pm}3/2}\right\rangle\}$. To the
lowest order the effective interaction in the low-energy subspace is of
Ising-type, as shown also in \cite{delgado2014emergence}, and thus does not
induce direct transitions between energy levels. Magnetic exchange between
two neighboring sites is governed by a third order effect, as reflected in
the effective coupling $J_{\mathrm{mem}}$ displayed in Eq.~\eqref{JeffSlow}.
This third order process is mediated by the virtual processes depicted in
Fig.~\ref{FigSlow} where two high-energy levels are populated.

Since the exchange is only a third order process in $J/D$, an eventual
magnetic transfer between neighboring sites would take place in the large
time scale of $1/J_{\mathrm{mem}}$. Those transfer mechanisms can thus be
ignored (for suitably large $D$) in all processes that are governed by lower
order mechanisms, such as the transfer in the higher energy subspace that we
will discuss later. Hence, it is legitimate to use this subspace as a memory
for storing information.

Notice that the difference between ferromagnetic and anti-ferromagnetic couplings, {\it i.e.} the sign of $J$,  does not alter the results of our
proposal, as the system is not initialized in the ground state.

\subsection*{Spin dynamics in the high-energy subspace}

\begin{figure}[t]
\centering
\includegraphics[width=.46\textwidth]{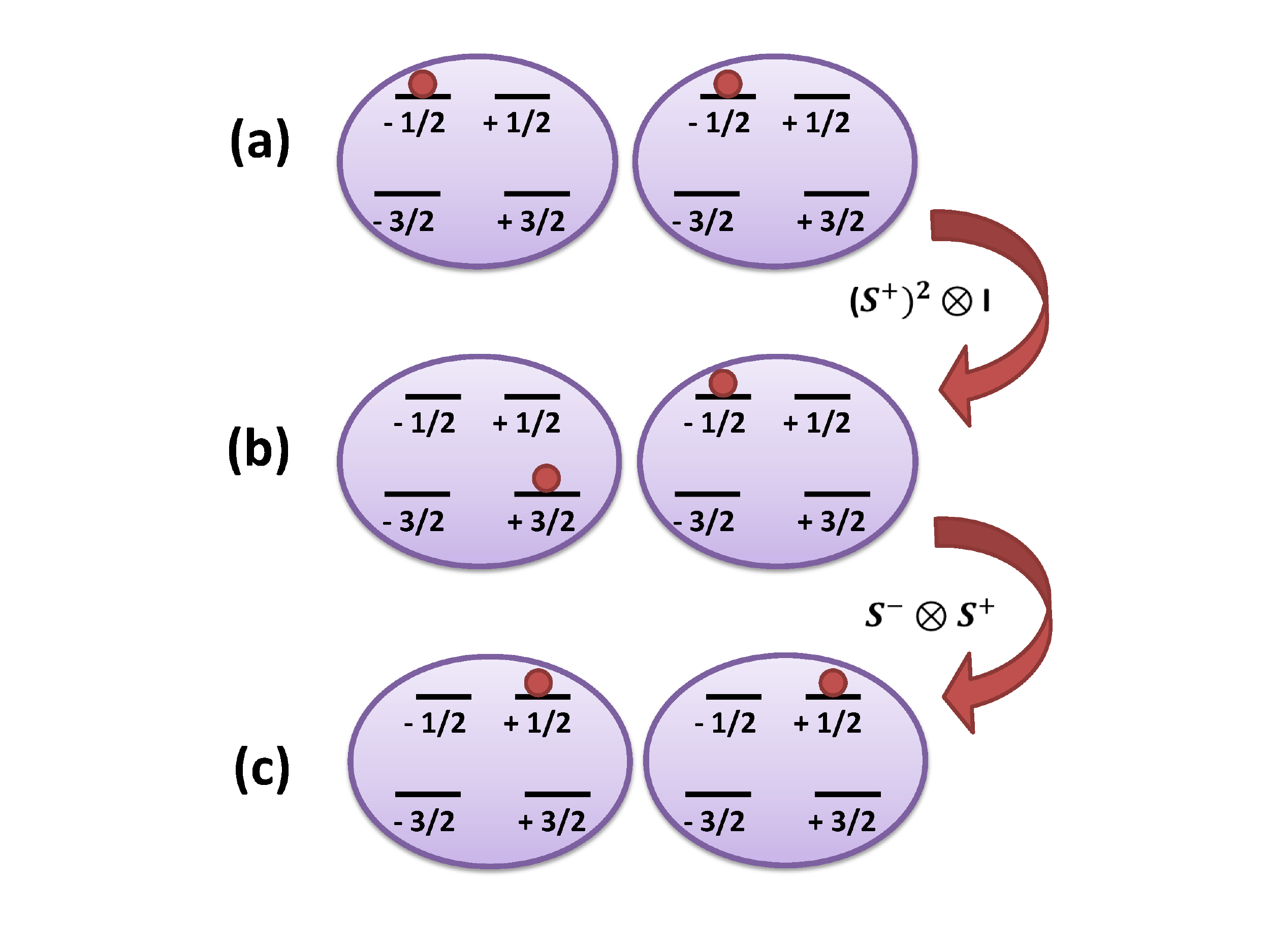}
\caption{ In the presence of in-plane anisotropy (i.e. $E>0$) the spin
exchange couplings $J_{\mathrm{bus}}^x$ and $J_{\mathrm{bus}}^y$ become
asymmetric in the $x$ and $y$ directions, as given in the effective
Hamiltonian of Eq.~\eqref{HeffFast}. The origin of this asymmetry is a
second order process through which the action of in-plane anisotropy $%
(S^+)^2 \otimes I$ (or its conjugate $(S^-)^2 \otimes I$) followed by the
operation of the usual spin exchange $S^+ \otimes S^-$ results in a term
like $\protect\sigma^+ \otimes \protect\sigma^+$ (or equivalently $\protect%
\sigma^- \otimes \protect\sigma^-$) in the effective Hamiltonian of the
high-energy subspace.
We show the states
(a) $\ket{\psi_a}=\ket{ {-}1/2,1/2}$,
(b) $\ket{\psi_b}=(S^+)^2\otimes I\ket{\psi_a}$,
(c) $\ket{\psi_c}=S^-\otimes S^+\ket{\psi_b}$,
which are populated during the third order process.
}
\label{ExplainAnisotr}
\end{figure}

We now consider the scenario in which the system is prepared in the high
energy subspace spanned by $\left|{{\pm}1/2}\right\rangle$. We again apply
the partial integration technique, described in the Supplementary Material. The resulting
effective Hamiltonian to the first order in $J/D$ is
\begin{eqnarray}  \label{HeffFast}
H_{\mathrm{bus}}^{\mathrm{eff}} = \sum_j &J^x_{\mathrm{bus}}\sigma^x_j
\sigma^x_{j+1} +J_{\mathrm{bus}}^y\sigma^y_j \sigma^y_{j+1} + \Delta_{%
\mathrm{bus}} \sigma_j^z\sigma_{j+1}^z +\nonumber \\
& \eta_{\mathrm{bus}} \left(\sigma_j^x\sigma_{j+2}^x
+\sigma_j^y\sigma_{j+2}^y\right) ~,
\end{eqnarray}
where the matrices $\sigma_{x,y,z}$ are the Pauli operators defined in the
effective subspace $\{ \left|{{\pm}1/2}\right\rangle\}$, and
\begin{align}
J_{\mathrm{bus}}^x &= J-\frac{3EJ}{D}~, & J_{\mathrm{bus}}^y
&= J+\frac{3EJ}{D}~, \\
\Delta_{\mathrm{bus}} &= \frac{J}4 - \frac{39J^2}{32 D}~, &
\eta_{\mathrm{bus}} &= - \frac{3J^2}{4 D}~.
\end{align}
As it is evident from the above formulae, $J_x=J_y=J$ to the zeroth order,
while there is a first order anisotropy in the $xy$ plane caused by the
crystal field anisotropy $E$. The origin of this effective anisotropy is
schematically explained in Fig.~\ref{ExplainAnisotr}
in which one state in the low energy subspace is virtually populated.
%

\begin{figure}[t]
\centering
\includegraphics[width=.48\textwidth]{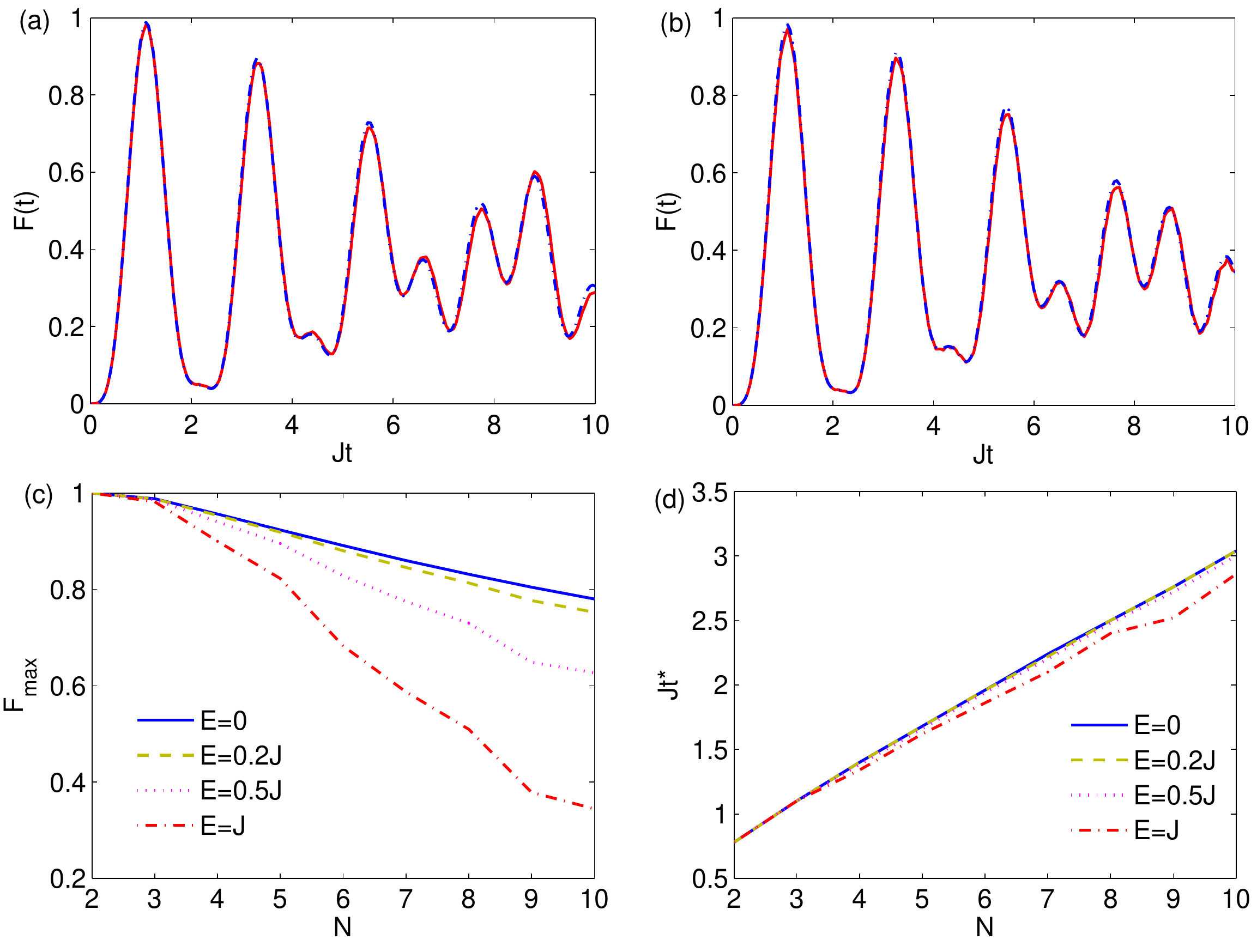}
\caption{(a),(b) Fidelity $F(t)$, given in Eq.~\eqref{Fid_tfast}, evaluated
with the real Hamiltonian (solid blue curve) and the effective Hamiltonian
(dashed red curve) acting on the data-bus subspace $\mathcal{H}_{\mathrm{bus}%
}$. The chain of length is $N=3$ and the parameters are $D=-20J$ and $E=0$
for Fig.\protect\ref{LongDistanceComm}a and $E=J$ for Fig.\protect\ref%
{LongDistanceComm}b. (c) The maximum of bit transfer fidelity $F_{max}$ as a
function of length $N$ for $D=-20J$ and different values of $E$.
(d) Scaling of the transfer
time $t^*$ as a function of the length $N$, using the same parameters of Fig.%
\protect\ref{LongDistanceComm}c. }
\label{LongDistanceComm}
\end{figure}

To see the performance of bit transfer in the data-bus subspace in a chain
of $N$ magnets we compute the fidelity
\begin{equation}
F(t)=\left\vert \left\langle {-}\frac{1}{2},...,{-}\frac{1}{2},\frac{1}{2}%
|e^{-iHt}|\frac{1}{2},{-}\frac{1}{2},...,{-}\frac{1}{2}\right\rangle
\right\vert ^{2}~.  \label{Fid_tfast}
\end{equation}%
We first calculate the fidelity with the Hamiltonian $H=
H_{\mathrm{tot}}$, to take account of the influence of the
lower-energy subspace on the information transfer in higher energy space.
We also compute the fidelity with $H=H_{\mathrm{bus}}^{\mathrm{eff}}
$ to check the validity of the effective Hamiltonian $H_{\mathrm{%
bus}}^{\mathrm{eff}}$ in higher-energy subspace. We make a
comparison of the time evolutions of the fidelity computed for the
total Hamiltonian $H_{\mathrm{tot}}$ and the effective Hamiltonian
$H_{\mathrm{bus}}^{\mathrm{eff}}$ for a spin chain of $N = 3$ with
the parameters $D=-20J$ and $E=0$ and $J$
respectively in Figs~\ref{LongDistanceComm}(a) and (b). The
perfect match of the two curves in Figs.\ref{LongDistanceComm}
(a) and (b) shows negligible influence of the lower-energy subspace
on transfer of the information initially written in the higher-energy
subspace along the nano-magnete chain, and thus the higher-energy subspace
could function as a data-bus for quantum information transfer.

In the high-energy subspace two neighboring spins are directly coupled by
the Heisenberg interaction \eqref{Htot}. This results in a very large
exchange coupling, (\textit{i.e.} $J_{\mathrm{bus}}\approx J$) which, in
turns, implies a very fast transfer dynamics compared with the low
energy subspace. As $J_{\mathrm{bus}}/J_{\mathrm{mem}}\approx (D/J)^{2}$,
one can see that in the regime where the effective Hamiltonian picture is
valid (say $D\geq 10J$), the high energy subspace is faster by at least two
orders of magnitude. This justifies the use of the high-energy subspace for
computational tasks with a fast bit transmission.


To quantify quality of the bit transfer one can consider the time $t=t^{\ast
}$ at which the fidelity $F(t)$, given in Eq.~\eqref{Fid_tfast}, peaks for
the first time and takes the value $F_{\mathrm{max}}=F(t^{\ast })$. In fact,
in real systems, decoherence deteriorates the quality of
bit-transfer and it is unwise to wait for later peaks. To see the
scalability of the bit transfer in larger chains we plot $F_{\mathrm{max}}$,
using the effective Hamiltonian \eqref{HeffFast}, as a function of length $N$
in Fig.~\ref{LongDistanceComm}(c). Although due to the non-linear dispersion
relation, the maximum fidelity $F_{\mathrm{max}}$ decreases with increasing $%
N$, it still remains above $0.75$ for $N\leq 10$.
{
  Moreover, in Fig.~\ref{LongDistanceComm}(c) we see that the transverse
  anisotropy $E$ has always a detrimental role for transmission, therefore
  those systems with vanishing $E$ are preferable.
  It is worth emphasizing that the effective Hamiltonian description is always
  valid for all $N$ provided that $|D|\gg J$. In view of this, the decrease of
  $F_{\rm max}$ as a function of $N$ is only due to the dispersive dynamics
  of the effective Hamiltonian (see {\it e.g.} \cite{initial}).
  In order to improve the fidelity for larger values of $N$ one has to
  linearize the dispersion relation through local engineering of
  the system parameters (such as the exchange coupling $J$),
  as extensively discussed in the literature \cite{kay_perfect_2010}.
}
As shown in Fig.~\ref%
{LongDistanceComm}(d), the optimal time $t^{\ast }$ linearly increases by $N$
but still is, at least, one order of magnitude faster than the time scale of
bit swap in the memory subspace, given by $1/J_{\mathrm{mem}}$.

Note that the distance over which a bit can be moved, strictly speacking,
needs not be limited. For example we can adapt some minimal engineering
techniques to move bits over much longer distances
\cite{banchi2011efficient}, though here the Hamiltonian is rather different.

\subsection*{Transitions between memory and data-bus subspaces}

We now study how one can simultaneously transfer all the states from the
low-energy to the high-energy subspace and \textit{vice versa}. When the
chain is composed of a single nanomagnet, the memory ($\{\left\vert {{\pm }%
3/2}\right\rangle \}$) and bus ($\{\left\vert {{\pm }1/2}\right\rangle \}$)
Hilbert subspaces have an energy separation of $2D$. A resonant spin
transition can be obtained with an electromagnetic pulse $\vec{B}(t)=B(\cos
(\omega t),0,0)$ with $\omega \simeq 2D$. This physically motivated
intuition can be made more rigorous. Indeed, near resonance, the
time-dependent interactions can be approximated in the rotating picture with
a time-independent Hamiltonian, and the off resonant energy levels are then
traced out using theory presented in the Supplementary Material. We found that in the
limit $\tilde{B}\equiv \mu _{B}gB\ll D$ a transition with $\Delta S^{z}=\pm
1 $ occurs with a transition time $\Delta t\simeq 2\pi /(\sqrt{3}\tilde{B})%
\left[ 1+\left( \tilde{B}/D\right) ^{2}/12\right] $ and a transition
fidelity $\simeq 1-\left( \tilde{B}/D\right) ^{2}/3$. Therefore, in the
limit $\mu _{B}gB\ll D$ the transition fidelity is almost one,
as it has been proved by our numerical simulations (not shown here).

On the other hand, in a many nanomagnet scenario the pulse has to be fast
enough to neglect the interaction $J$ between neighboring magnets, so the
optimal working regime is $J\ll \mu_B gB\ll D$.

\subsection*{Higher spin systems}

The procedure described above can be applied also for higher half-integer
spin $S$ systems. In fact, the dynamics in the high energy subspace $\left|{{%
\pm}1/2}\right\rangle$ is still governed by an effective Hamiltonian which
has a similar form of Eq.~\eqref{HeffFast}. In particular, the leading term
is an exchange coupling which results in fast transmission times ($\approx
1/J$).
In contrast, the storage quality of the memory subspace
$\left|{{\pm}S}\right\rangle$ is significantly improved as $S$ increases.
Indeed, the effective coupling $J_{\mathrm{mem}}\propto J (J/D)^{(2S-1)}$
becomes smaller by increasing $S$, making the storage much less prone to
errors over a longer time scale.
{
However, in some higher-spin systems, like rare-earth ions, there might
be higher order anisotropy terms (i.e. Stevens operators) in the Hamiltonian
\cite{stevens,stevens1952matrix} which might change the effective couplings.
}

In addition, bit-flip errors, i.e. spurious transitions between the states
$\left|{{\pm}S}\right\rangle$, are exponentially suppressed as they
require multiple jumps through higher energy
levels. 
This, however, comes with the price that a transition from the memory to the
data-bus subspaces demands multiple pulse sequences (namely $(2S-1)/2$
consecutive pulses) which increase the complexity of the process.

\section*{Imperfections}
In this section we consider two sources of imperfections which may
affect our protocol, namely decoherence and possible
long-range interactions arising, {\it e.g.}, from dipolar couplings between
distant nanomagnets.

\subsection*{Effect of decoherence}

In a real physical system it is notoriously difficult to keep the system
isolated from the surrounding environment. Depending on the nature of the
interaction between the system and the environment one may have different
decoherence processes. In particular, for the main target experiments of our
theoretical proposal, e.g. molecular magnets and magnetic adatoms on
surfaces, the dissipation time $T_{1}$ is larger than the dephasing time $%
T_{2}$ by several order of magnitudes. Dephasing, which is thus the main
source of decoherence, arises because of complicated interactions with other
degrees of freedom. In this paper we consider a simple model of decoherence,
i.e. caused by random energy level fluctuations due to nearby magnetic and
electric impurities. By averaging over the possible random time fluctuations
one obtains a master equation for the evolution of the system which has the
Lindblad form \cite{delgado2012storage}%
\begin{equation}
\dot{\rho}=-i[H,\rho ]+\gamma \sum_{j}\left( L_{j}\rho L_{j}^{\dagger }-%
\frac{1}{2}L_{j}^{\dagger }L_{j}\rho -\frac{1}{2}\rho L_{j}^{\dagger
}L_{j}\right) ~,
\label{lindblad}
\end{equation}%
where $\rho (t)$ is the system density matrix, $\gamma $ is the dephasing
rate and $L_{j}$ are Lindblad operators. Due to the fact that $T_{1}\gg T_{2}$,
we only consider dephasing here, which can be modeled with $L_{j}=S_{j}^{z}$%
.
Although the master equation \eqref{lindblad} neglects non-Markovian effects, it
is widely used to model qualitatively the action of the environment
in the type of systems that we consider for physical realization
\cite{delgado2012storage}.

We first consider the case where the system is prepared in the high-energy
subspace for computational tasks and we study the effect of dephasing on the
fidelity of state transmission. As an example the two-site system is
initialized in the pure state $\rho (0)=\left\vert {\psi _{\mathrm{init}}}%
\right\rangle \langle {\psi _{\mathrm{init}}}|$, where $\left\vert {\psi _{%
\mathrm{init}}}\right\rangle =\left\vert {1/2,-1/2}\right\rangle $, but
because of the non-unitary evolution \eqref{lindblad} it evolves into a mixed
state $\rho (t)$. The resulting fidelity of state swap is therefore $%
F(t)=\langle {{-}1/2,1/2}|\rho (t)\left\vert {{-}1/2,1/2}\right\rangle $.
\begin{figure}[t]
\centering
\includegraphics[width=.4\textwidth]{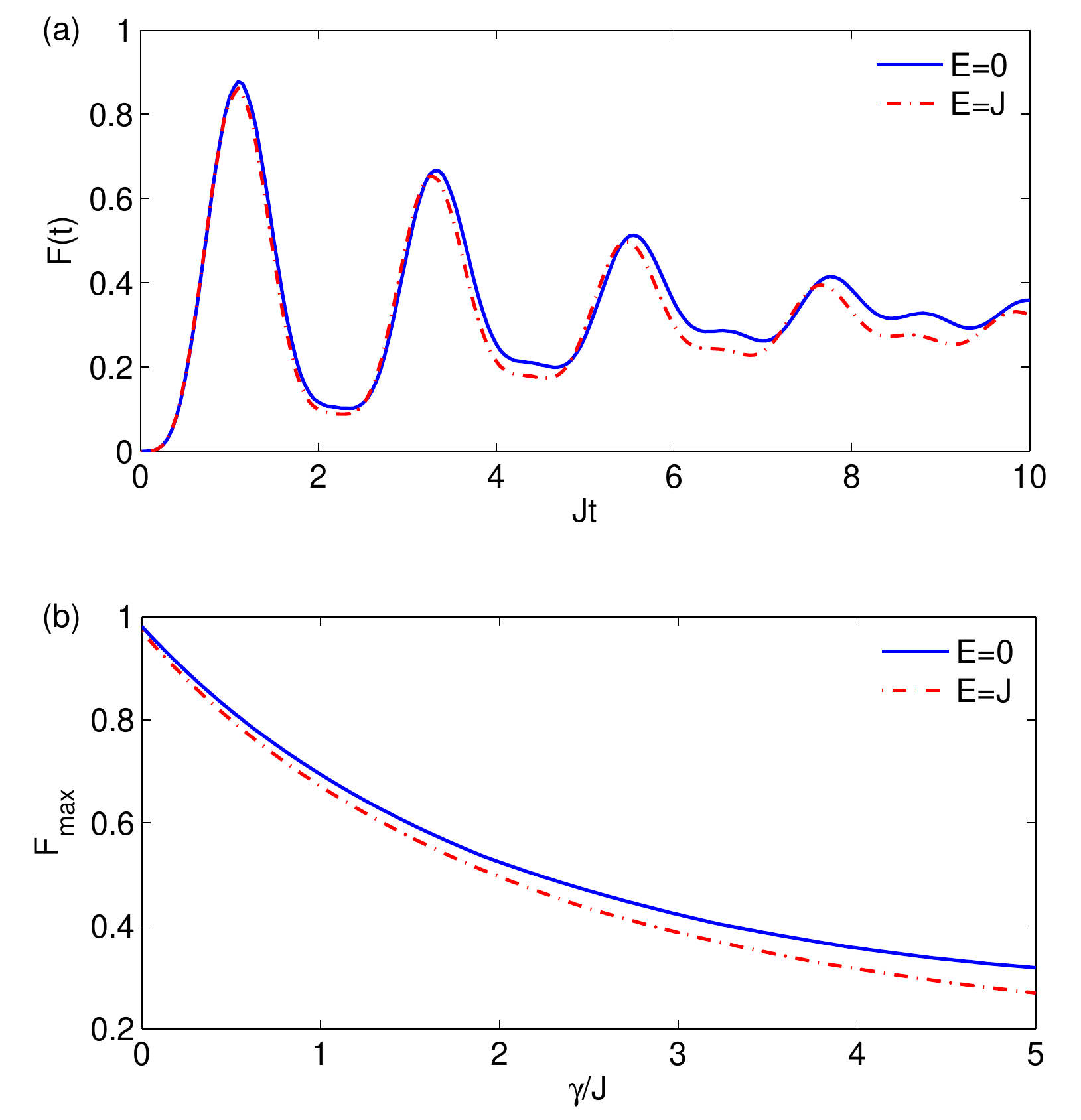}
\caption{(a) The bit-tranfer fidelity $F(t)$
  in the data-bus subspace with dephasing rate $\gamma =0.5J$ when
  $D=-20J$ and $N=2$. The chosen value for $\gamma$ is extremely  pessimistic
  even for larger chains, and we have chosen this value in order to show
  the decay in shorter time scales. (b) The maximum fidelity $F_{max}$ as a
function of $\protect\gamma $ for two different values of anisotropy $E$,
when $D=-20J$ and $N=2$.}
\label{DephasingFast}
\end{figure}
In Fig.~\ref{DephasingFast}(a) we plot $F(t)$ as a function of time for
a very strong $%
\gamma =0.5J$.
We have chosen a high value of $\gamma$ to show its effect on the coherent
dynamics of our system within shorter time scales. The realistic values
are indeed much smaller ($\gamma\approx 10^{-3}J$ as discussed in the
next section) and allows for very high quality transfer even in long chains.
Due to the damping dynamics shown in the plot it is wise to
only concentrate on the first peak of the fidelity $F_{\mathrm{max}}$. The
latter quantity is displayed in Fig. \ref{DephasingFast}(b) as a function of
the dephasing rate $\gamma $ and for different values of $E$. As it is
expected $F_{\mathrm{max}}$ exponentially decays with the increase of
 $\gamma $. The decay rate only weakly depends on $E$ and slightly become
faster for larger $E$.
\begin{figure}[t]
\centering
\includegraphics[width=.41\textwidth]{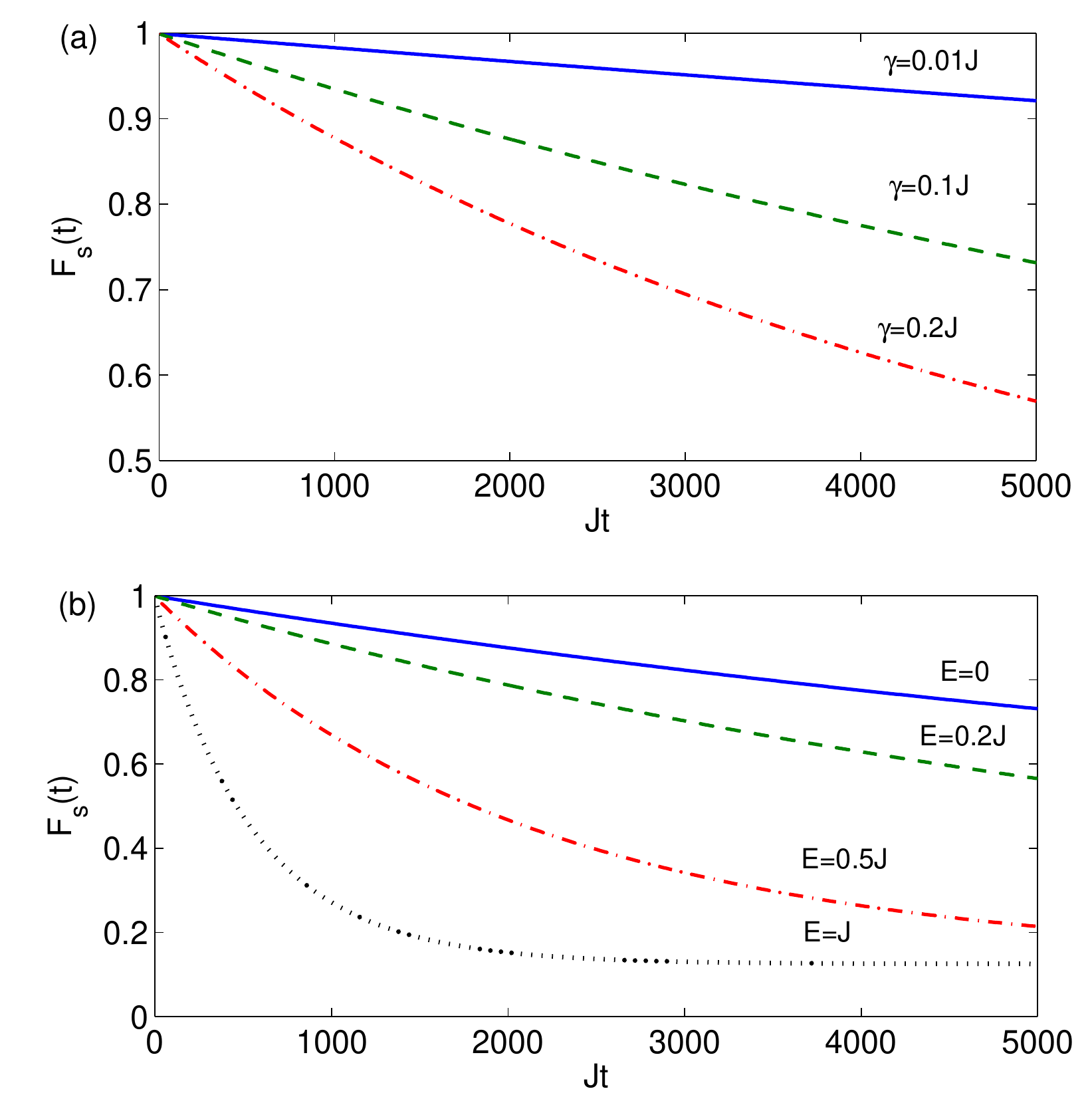}
\caption{Storage fidelity $F_{s}(t)$ in the memory subspace as a function of
time for: (a) Different dephasing rates $\protect\gamma $ and; (b) Different
in-plane anisotropy $E$. In both figures $N=2$ and $D=-20J$.}
\label{DephasingSlow}
\end{figure}

We now study the effect of dephasing on information storage, namely when the
system is prepared in the low-energy subspace. To investigate the quality of
the storage we define a new fidelity which measures the deviation from the
initial state at any time $t$. For example, we consider the initial pure
state $\left\vert {\psi _{\mathrm{init}}}\right\rangle =\left\vert {{-}%
3/2,3/2}\right\rangle $ and we define the storage fidelity as $F_{\mathrm{s}%
}(t)=\langle {\psi _{\mathrm{init}}}|\rho (t)\left\vert {\psi _{\mathrm{init}%
}}\right\rangle $ where $\rho (t)$ is calculated from the master equation %
\eqref{lindblad}. In Fig.~\ref{DephasingSlow}(a) we study the
time evolution of the $F_{s}(t)$ for different values of $\gamma $, when $%
E=0$. As expected the storage fidelity decays in time with a rate which
increases for increasing $\gamma $. However, within the timescale of tens of
operations in the computational subspace (say $Jt\approx 100$) the quality
of the storage is only weakly affected by dephasing, as $F_{\mathrm{s}}$
remains above 0.95 even for a strong dephasing of $\gamma =0.2J$.

Unlike the high-energy subspace in which the transmission was hardly
affected by the in-plane anisotropy $E$, in the storage subspace the effect
is no longer negligible after long times. To show this effect we plot in
Fig.~\ref{DephasingSlow}(b) the storage fidelity as a function of time for
different values of $E$ when $\gamma =0.5J$. It turns out that the storage
fidelity decays faster for larger values of $E$, due to enhanced
coupling of the low and high energy subspaces by the in-plane anisotropy.
Since $E$ depends on the geometric property of the material, it is
preferable to use systems with vanishing $E$ to guarantee longer
storage times.

It is worth emphasizing that the stability of the low energy subspace with
respect to dephasing is significantly enhanced by using larger spins because
the effective coupling between neighboring sites exponentially goes down by
increasing the spin $S$.

\subsection*{Long-range interactions}
The dipolar or RKKY coupling between nanomagnets may induce interactions beyond
the nearest neighbours.
To study this effect we modify our Hamiltonian $H_{\rm tot}$ in \eqref{Htot}
such that two nano-magnets at sites $i$ and $j$ interact as
$J_{ij}\vec S_i \cdot \vec S_j$, where $J_{ij}{=}J/|i-j|^{3}$.

\begin{table}
  \centering
  \begin{tabular}{|c|c|c|c|c|c|c|c|c|c|}
          \hline
          N  &  2 &      3&       4&       5&      6&       7&        8&
          9 &      10\\
          \hline
          $F^{\rm max}_{\rm SR} $ &  0.9993 &  0.9829&  0.9464&  0.9035
          &0.8764  &0.8409& 0.8005&  0.7776 & 0.7535  \\
          \hline
          $Jt^*_{\rm SR}$ & 0.79  &   1.12 &  1.38  &   1.73 &  1.98 &
        2.23& 2.47 &   2.83 &    3.07  \\
          \hline
        $F^{\rm max}_{\rm LR}$ &  0.9993 &  0.9805 & 0.9512&  0.9105 &
      0.8653 & 0.8339 &0.8058 & 0.7753 &0.7438\\
          \hline
      $Jt^*_{\rm LR}$ &   0.79    & 1.10  & 1.37   &  1.62  & 1.86  &
    2.20 &2.43 &2.66 &2.90\\
    \hline
  \end{tabular}
  \caption{Maximum bit transfer fidelity $F^{\rm max}$ and its corresponding
  transfer time $t^*$ as a function of $N$
  for nearest neighbour, namely short-range (SR), interactions
  \eqref{Htot} and long-range (LR) interactions.
}
  \label{tab:longrange}
\end{table}

In Table~\ref{tab:longrange} we show the maximum transfer fidelity $F^{\rm max}$, and its
corresponding transfer time $t^*$,
in terms of length $N$ for both nearest neighbour
and long-range interactions.
As it is clear from Table~\ref{tab:longrange}
the long-range interaction has little effect on the information
transfer along the spin chain.

\section*{Implementation}
\subsection*{Molecular magnets}

We propose an array of molecular magnets for realizing our theoretical
proposal.
Indeed, single-molecule magnets are very attractive because of
many reasons: (i) they can be prepared by chemical synthesis in a huge range
of configurations providing scalability for quantum technology \cite%
{sanvito2011molecular}; (ii) they are composed of spin clusters and can be
individually addressed because of their large size \cite{bogani2008molecular}%
; (iii) very small ratios of $J/D$ ($\approx 10^{-2}$) \cite%
{timco2009engineering} have already been realized; (iv) the dissipation time $%
T_{1}$ is extremely large ($\approx 4\mathrm{ms}$) \cite{luis2011molecular},
and the decoherence time $T_{2}$ exceeds few $\mu \mathrm{s}$ \cite%
{ardavan2007will,mitrikas2008electron}; (v) the in-plane anisotropy $E$
is negligibly small \cite{timco2009engineering}.
  Molecular rings,  such as heterometallic wheels Cr$_7$M
  \cite{ardavan2007will} (M is a metal center),
  are promising candidates for realizing our proposed mechanism.
  By using different metal centers M, one can change the spin sector of the
  ground state:
  for instance, $S=3/2$ is achieved with M=Ru$^{2+}$Ru$^{3+}$
   \cite{timco2009engineering}.
  The entanglement properties
  \cite{siloi2013quantum,siloi2014detection}
  of such rings have been studied and there are proposals to split them
  into open chains \cite{apollaro2013effective}.

Local addressability can be achieved by engineering the $g$-factor in
different sites through chemical synthesis \cite%
{Nakahara:1320438,nakazawa2012synthetic,troiani2011molecular}. This indeed
creates site dependent Zeeman energy splitting in the presence of a uniform
magnetic field, even without the complexity of a spatially modulated field.
Such engineered $g$-factors allows initialization and readout of specific
sites using selective microwave pulses which are in resonance only with the
target site and practically have no effect on the rest. While the magnetic
field is needed for local addressability, as a part of initialization and
readout, it should be switched off otherwise.

Initially the system can be prepared in the ferromagnetic state where all
the magnets are aligned in the same low-energy quantum state $\left|{S^z{=}{-%
} S}\right\rangle$ by applying strong magnetic fields \cite%
{bogani2008molecular,gauyacq2013magnetic}.
To write the information in the memory subspace the magnetization of each
site can be selectively reversed $\left|{-S}\right\rangle\to\left|{+S}%
\right\rangle$ by applying a fast sequence of electromagnetic pulses \cite%
{choi2012coherent} or a suitably modulated multi-frequency pulse \cite%
{leuenberger2001quantum}.

For most of the time, the molecules stay in the low energy subspace, which
effectively do not evolve. To transfer information between distant sites one
needs to bring all the molecules into the high-energy subspace $\left|{S^z{=}%
{\pm}1/2}\right\rangle$ in which the interaction between neighboring sites
is strong. Such transitions can be implemented via global electromagnetic
pulses which act collectively on all magnets simultaneously. Each pulse
makes a magnetic transition with $\Delta S^z=\pm 1$, till the state is
transfered to the high-energy subspace $S^z=\pm1/2$. For instance, for $%
S=3/2 $ such transition is achieved in a single step. An alternative
approach is via using a properly modulated multi-frequency pulse which makes
these transitions in a single step \cite{leuenberger2001quantum}.
After
finishing the transfer, the same set of pulses can be used for bringing
the states back to the memory subspace.
Since the $g$-factor is site-dependent, as required for local addressability,
the transition time for each site will be different. In order to achieve
the transition from memory to data-bus subspace
(and {\it vice versa}) with a single global operation,
one may use an
adiabatic inversion pulse
\cite{Kupce1995273}
which is intense and operates within a short period of time.
Consequently, this pulse
has a wider frequency spectrum
capable of exciting all sites in spite of
the different resonance energies.
  There are various ways of implementing such pulses
  \cite{KupceNew}, each one with its specific duration and intensity.
  For instance, using a linear frequency sweep with range $\Delta f$,
  one requires a pulse duration $\propto \Delta f/{\tilde B}^2$
  \cite{KupceNew}.

Finally, when the system is back into the memory subspace, thanks to the
slow dynamics of the low-energy subspace, there is enough time for readout.
Magnetic readout has been experimentally realized with different ways,
either with a scanning tunneling microscope (STM) \cite{sanvito2011molecular}
or with electronic paramagnetic resonance (EPR) \cite%
{luis2011molecular,timco2009engineering,choi2012coherent}.

Molecular magnets represent a flexible setup as their magnetic properties
can be engineered in a wide range via chemical synthesis. Promising
molecules for quantum information applications \cite%
{timco2009engineering,luis2011molecular} display a small value of $J/D$ and
a large decoherence time. For instance,
  using heterometallic wheels Cr$_7$M
\cite{timco2009engineering},
the values $J\approx 100\mathrm{GHz}$
and $D\approx 88J$ have been measured.

A typical \cite{luis2011molecular} $J=2\mathrm{GHz}$ and $D=-20J$
implies a transmission time of $\approx 1\mathrm{ns}$. Due to the very
large $T_1$ (e.g. $\approx 1 \mathrm{ms}$ in \cite{luis2011molecular}) the
limiting time scale is given by $T_2$ which exceeds $1 \mu\mathrm{s}$ \cite%
{timco2009engineering,ardavan2007will,mitrikas2008electron}.
In our formalism this corresponds to $\gamma\approx 1 {\rm MHz}$ and
therefore $\gamma/J \approx 10^{-3}$.
This allows for
$\approx 10^3$ operations before equilibration. The technology allowing very
fast pulse sequences has already been developed \cite{choi2012coherent}
paving the way for controlling the dynamics in the sub-nanosecond regime.
This opens the possibility of using molecular clusters with larger exchange
interaction $J$ allowing even more operations within the coherence time.

\subsection*{Adatoms on surfaces}
Another exciting possibility are structures made from magnetic adatoms
(e.g., Co, Mn etc.) created and probed on surfaces using STM \cite%
{hirjibehedin2013spintronics}. Recently adatomic clusters \cite%
{loth2012bistability}, and adatoms themselves \cite{delgado2012storage} on
surfaces have indeed been proposed as a quantum storage of a classical bit.
Their dephasing has been studied using the same type of weak coupling
Lindbladian master equations as considered by us here \cite%
{delgado2012storage}. This approach has become even more accurate very
recently with the advent of superconducting layers replacing the usual two
dimensional electron gas in STM so as to greatly increase electron
relaxation times for the adatoms \cite{heinrich2013protection}.
As electron
spin scattering is suppressed because of the energy gap of the
superconductor, we also naturally expect the dephasing time to be enhanced
in addition to the relaxation time. Though the dephasing time is yet to be
measured, this kind of work is ultimately aimed towards taking adatoms
towards the regime of coherently operating devices.
Another technique by which the effective isolation of adatoms has been
greatly enhanced is by using symmetry protected systems
\cite{miyamachi2013stabilizing,flatte2013quantum}.
These give the hope that adatoms will eventually approach the coherent
regime \cite{flatte2013quantum} so that coherent non-equilibrium
dynamics, as used in our paper, will become accessible.
Very large anisotropies $%
D$ have also been recently achieved \cite{rau2014reaching} for Co atoms, for
example. Microwaves could still be used for changing between the memory and
data-bus modes of the chain, but measurements can be done locally at leisure
using spin polarized STMs \cite{khajetoorians2011realizing} after setting
the device to memory mode. Moreover, we can bring a magnetic tip close to
the adatom (as in the newly devised magnetic exchange force microscopy, which
is compatible with STM setup)
to apply a local field to it \cite{wieser2013theoretical}.
This local field, if in a pertinent direction, can directly precess the
adatom's spins. Alternatively, it can locally Zeeman split the energy levels
so that a microwave can locally flip it.
In particular, a more
macroscopic nanomagnetic bit attached to the STM tip could be made to talk
to the adatom bit by bringing the tip in proximity to it
\cite{kong1997writing}. This may offer a
route to interface the system we discussed here with more conventional
magnetic memory with larger magnetic bits.

\section*{Conclusions}


A general problem in any information processing architecture is that memory
cells are supposed to be well decoupled from each other to act as a good
information storage, while the data-bus cells should have strong
interactions to implement fast quantum gates and information transfer. The
different interactions required in the memory and data-bus units make
it very challenging to spot a physical system suitable for both, so one way
to face this problem is to use hybrid structures \cite%
{monroe2014large,xiang2013hybrid,carretta2013quantum} which demands very
sophisticated control of the system. To avoid such complexities, in this paper
we have proposed a mechanism to implement memory and data-bus units,
two key requirements for any processor, with the same physical setup,
namely arrays of large-spin magnets.
The data-bus and memory subspaces are encoded into different spin levels of the
same magnet. The selection rules imposed by the exchange coupling together
with a large zero field energy splitting result in different effective
coupling between spin levels of the two neighboring magnets. While
high-energy spin levels are directly coupled by the exchange interaction,
lower energy subspaces are coupled only via higher-order processes, which
become ineffective in the time scales of operations in the high-energy
subspace. Hence, the high-energy subspace is suitable for transmission
tasks, while the low-energy subspace can act as a robust memory. Transitions
between the two subspaces can be done at will by applying resonant external
pulses.

Despite having not gone into the details of how one would engineer
gates between the magnetic bits, we would like to point out
that bit movement is already a significant step towards it.
For example, for other (incoherent) mechanisms of bit movement, gates
were immediately accomplished by bringing two bits in close proximity
\cite{khajetoorians2011realizing}.
Of course, in continuation with our coherent bit movement protocol,
we would expect that a XOR gate should be implementable for
bits brought next to each other (by data-buses) and in memory subspace
through their dominantly Ising interaction of Eq.\eqref{HeffSlow}
\cite{nielsen2010quantum}.

The theory that has been developed does not depend on a particular physical
realization and can be applied to many systems, such as magnetic adatoms on
surfaces \cite{rau2014reaching}, donors on silicon \cite%
{kane1998silicon,zwanenburg2013silicon} and molecular magnets \cite%
{troiani2011molecular,timco2009engineering,Nakahara:1320438,ardavan2007will,mitrikas2008electron}%
.  The desirable requirements for our scheme are:
(i) addressability of some individual
magnets to accomplish read/write operations;
(ii) long coherence times for transfer over tens of magnetic
cells, but not as demanding as for quantum computation; (iii)
flexibility in engineering the couplings; (iv) large zero-field energy
splitting; and (v) intrinsically vanishing in-plain anisotropy. Our proposed
mechanism can be realized in molecular magnets
with current technology, and we showed that, using
parameters taken from recent experiments, it allows for hundreds of
operations within the coherence time of the system.

\section*{Acknowledgements}

We warmly acknowledge discussions with
L. Sorace, C. F. Hirjibehedin, and A. A. Khajetoorians.
LB and SB acknowledge the support of the ERC grant \textquotedblleft
PACOMANEDIA\textquotedblright . AB is supported by the EPSRC grant
EP/K004077/1. Guangjiong Dong achnowledges the support of the National Basic
Research Program of China ("973" Program, No. 2011CB921602), the National
Natural Science Foundation of China (No. 11034002), and the Program of
Introducing Talents of Discipline to Universities (B12024), as well as the
Research Fund for the Doctoral Program of Higher Education of China (No.
20120076110010).

\section*{Author contributions}
SB proposed the original idea.
NJ carried out all numerical simulations. NJ, LB, and GD derived the analytical
effective Hamiltonian theory. LB, AB and SB developed the experimental
proposal. All the authors
contributed to prepare the final version of the manuscript.

\section*{Additional information}
{\bf Competing financial interests}: The authors declare no competing financial
interests.

\end{document}